\documentclass[prb,
twocolumn,
superscriptaddress,showpacs,amsmath,amssymb]{revtex4}

\begin{document}

\title{Macroscopic quantum tunneling: from quantum vortices to black holes and Universe}

\author{G.E.~Volovik}
\affiliation{Low Temperature Laboratory, Aalto University,  P.O. Box 15100, FI-00076 Aalto, Finland}
\affiliation{Landau Institute for Theoretical Physics, acad. Semyonov av., 1a, 142432,
Chernogolovka, Russia}

\date{\today}

\begin{abstract}
The paper has been prepared for the JETP issue, dedicated to the 95th anniversary of the birth of E.I. Rashba.
E. Rashba stood at the origins of macroscopic quantum tunneling together with his colleagues from the Landau Institute S.V. Iordansky and A.M. Finkelshtein. They pave the way for studying macroscopic quantum tunneling in various systems. In this paper, macroscopic quantum tunneling approach is extended to cosmological objects such as a black hole and de Sitter Universe. In particular, this approach allowed to calculate the  entropy of Reissner-Nordstr\"om (RN) black hole with two horizons and the corresponding temperature of the thermal Hawking radiation.
Several different methods were used: the method of semiclassical tunneling for calculation of the Hawking temperature; 
the cotunneling  mechanism -- the coherent sequence of tunneling at two horizons, each determined by the corresponding Hawking temperature; the method of singular coordinate transformations for calculations of macroscopic quantum tunneling from the RN black hole to the RN white hole; the method of the adiabatic change of the fine structure constant for the adiabatic transformation from the RN black hole to the Schwarzschild black hole; etc. 
\end{abstract}
\pacs{
}

\maketitle

 \tableofcontents


\section{Introduction}

The process of quantum tunneling of macroscopic objects is well known in condensed matter physics, where the collective variables are used, which describe the collective dynamics of a macroscopic body. \cite{LifshitzKagan1972,IordanskiiFinkelshtein1972,IordanskiiRashba1978}
This approach allows to estimate the semiclassical tunneling exponent without consideration of the details of the structure of the object on the microscopic (atomic) level.

\subsection{Quantized vortex vs black hole}

One of the application of the macroscopic quantum tunneling is the calculation of the quantum creation of the topological objects. Examples are the nucleation quantized vortices in moving superfluids;\cite{Volovik1972}  nucleation of Abrikosov vortices in superconductors in the presence of supercurrent;\cite{Blatter1994} and the instanton -- the process of creation of the topological charge in quantum field theories.\cite{BPST1975} Example is the vortex instanton, which presents the process of quantum nucleation of the vortex ring.\cite{Volovik1972} The corresponding collective (macroscopic) dynamically conjugate variables are the area $A=\pi R^2$ of the created vortex ring with radius $R$ and (with some factor) its coordinate $z$ along the normal to the ring. 

It looks reasonable to apply the approach of  macroscopic quantum tunneling also to such macroscopic objects as a black hole. In this case the corresponding  collective variables\cite{Volovik2020a} are the area of the event horizon $A=4\pi R^2$ and its dynamically conjugate variable -- the gravitational coupling $K$ (we use the gravitational  coupling $K=1/(4G)$, where $G$ is  the Newton "constant").

Since in both cases one of the collective variables is represented by the corresponding area, this suggests that there can be some thermodynamic analogy between the vortex ring and the black hole. And indeed it was shown in Ref. \cite{Volovik1995} that quantized vortices in Fermi superfluids have many common properties with the black holes.
 In particular, there is analog of the Hawking temperature for the moving vortex ring, see Eq.(4.1.9) in Ref. \cite{Volovik1995}:
 \begin{equation}
T_{\rm H}= \frac{\hbar v_F}{4\pi R} \ln \frac{R}{r_c}  \,,
\label{TVortexRing}
\end{equation}
where $v_F $ is Fermi velocity, and $r_c$ is the radius of the singularity -- the vortex core radius. In Fermi supefluids, the core size is the analog of the Planck length, which determines singularity inside the black hole.
In the 2+1 dimensional case, where the counterpart of the vortex loop is the pair of oppositely
oriented point vortices, there is no singularity:
\begin{equation}
T_{\rm H}= \frac{\hbar v_F}{\pi R}   \,,
\label{TVortexPair}
\end{equation}
where $R$ now is the distance between the point vortices.

The temperature in Eqs. (\ref{TVortexRing}) and  (\ref{TVortexPair}) looks similar to the Hawking temperature of black holes:
 \begin{equation}
T_{\rm H}= \frac{\hbar c}{4\pi R}   \,.
\label{Tbh}
\end{equation}
The analogy with black holes is supported by the behavior of fermionic quasiparticles living in the vortex core.  They occupy the  bound states -- the Caroli-de Gennes-Matricon states.\cite{CarolideGennesMatricon1964} Due to the motion of the vortex ring the fermions  are excited from the bound states to the continuous spectrum by the process of quantum tunneling. The tunneling exponent reproduces the thermal nucleation with the analog of Hawking temperature in Eq.  (\ref{TVortexRing}). If to extend this analogy to the black hole, then the Hawking radiation from the black hole can be considered as the quantum tunneling  of particles from the bound state inside the black hole singularity to the continuous spectrum outside the event horizon. The Hawking radiation as semiclassical tunneling was considered in Refs. \cite{Wilczek2000,Srinivasan1999,Volovik1999,Akhmedov2006,Vanzo2011} and in many following papers.

The analogy between the vortex rings in Fermi superfluids (with their fermion zero modes in the vortex core) and the black holes concerns both the individual microscopic processes of the particle creation by quantum tunneling from the object to the external world, and the related macroscopic processes of  quantum tunneling of the whole objects. In this paper we discuss the processes of microscopic and macroscopic quantum tunneling related to the black hole and de Sitter Universe using experience with the objects in condensed matter, where we know physics both on macro and micro scales. 
The thermodynamics of black holes and of the cosmological horizon is still confusing.\cite{Maldacena2021} In this paper we are trying to clarify some issues using both microscopic and macroscopic processes of quantum tunneling.

The plan of the paper is the following.

\subsection{Macroscopic black hole tunneling}

In Sec. \ref{KA} we consider the macroscopic   quantum tunneling of Schwarzschild black hole (the spherical electrically neutral black hole) to the Schwarzschild white hole of the same mass using   inverse Newton constant $K=1/(4G)$ as dynamic and thermodynamic variable. Introduction of the variying $K$  modifies the first law for the Schwarzschild black hole thermodynamics: $dS_\text{BH}= -AdK + \frac{dM}{T_\text{BH}}$, where $M$ is the black hole mass, $A=4\pi R^2$ is the area of horizon, and $T_\text{BH}$ is Hawking temperature in Eq.(\ref{Tbh}).
From this first law it follows that the dimensionless quantity  $M^2/K$ is  the adiabatic invariant, which in principle can be quantized if to follow the Bekenstein conjecture.\cite{Bekenstein1974} 

From the Euclidean action for the black hole it follows that $K$ and $A$ serve as dynamically  conjugate variables. Using the Painleve-Gullstrand metric, which in condensed matter is known as acoustic metric,  we calculated the quantum tunneling rate from the black hole to the white hole. The obtained tunneling exponent -- the probability of the macroscopic quantum tunneling is $p \propto \exp(-2S_\text{BH})$, where $S_\text{BH}$ is the entropy  of the Schwarzschild black hole. The tunneling transition can be considered as the random fluctuation, which probability is determined by the difference in entropy before and after tunnelling. Then  $p \propto \exp(S_\text{WH}-S_\text{BH})$, and comparing this with $p \propto \exp(-2S_\text{BH})$ one obtains that  the entropy of the white hole is with minus sign the entropy of the black hole, $S_\text{WH}(M)=- S_\text{BH}(M)= - A/(4G)$.  The temperature of the white hole is also negative.

\subsection{Singular coordinate transformations and macroscopic quantum tunneling}

In Sec. \ref{coordinate} we use another method of calculations. We consider three different types of the hole objects: the black hole, the white hole and the fully static intermediate state. The probability of tunneling transitions between these three macroscopic oblects is found using singularities in the coordinate transformations between these states. The black and white holes are described by the Painleve-Gullstrand coordinates with opposite shift vectors, while the intermediate state is described by the static Schwarzschild coordinates. 
The singularities in the coordinate transformations lead to the imaginary part in the action, which determines the tunneling exponent. For the white hole the same negative entropy is obtained, $S_\text{WH}(M)=- S_\text{BH}(M)= - A/(4G)$, while the intermediate state -- the fully static hole --  has zero entropy. 

\subsection{Macroscopic quantum tunneling and entropy for black holes with two horizons}

This procedure is extended to the black hole with two horizons in Sec.\ref{TwoHorizons}. We consider the electrically charged black hole -- the Reissner-Nordstr\"om black hole. We present the calculations of  the quantum tunneling of the Reissner-Nordstr\"om black hole to its white and static partners,  the  entropy of Reissner-Nordstr\"om black hole and the corresponding temperature of the thermal Hawking radiation. We use several different approaches, but all of them give the same result: the entropy and temperature of Hawking radiation depend only on mass $M$ of the black hole and do not depend on the black hole charge $Q$. This is the consequence of the correlations between the inner and outer horizons of the Reissner-Nordstr\"om black hole.

\subsection{de Sitter Universe: quantum tunneling and thermodynamics}

In Sec. \ref{dS} the consideration is extended to the entropy and temperature of the expanding de Sitter Universe.
We show that as distinct  from the black hole physics, the de Sitter thermodynamics is not determined by the cosmological horizon.   While the formal semiclassical calculations of the  tunneling rate across the cosmological horizon is described by the Hawking temperature $T_{\rm Hawking}=H/2\pi$, the effective temperature of the de Sitter Universe differs from $T_{\rm Hawking}$. 

In particular, atoms in the de Sitter Universe experience thermal activation corresponding to the local temperature, which is twice larger than the Hawking temperature,  $T_{\rm loc}=2T_{\rm Hawking}=H/\pi$. The same double Hawking temperature describes the decay of massive scalar field in the de Sitter universe. We show that the reason, why the local temperature is exactly twice the Hawking temperature, follows from the geometry of the de Sitter spacetime. The weakening of the role of the cosmological horizon in de Sitter universe is confirmed by the proper consideration of the Hawking radiation and of the macroscopic quantum tunneling.

\section{Macroscopic tunneling from black hole to white hole using the collective variables}
\label{KA}

\subsection{Collective variables of Schwarzschild black hole}
\label{CollectiveVar}

For macroscopic description of the macroscopic quantum tunneling we need two collective variables which are relevant for the black hole. 
One of them is  the horizon area $A=4\pi R^2$. Bekenstein \cite{Bekenstein1974}  proposed that  $A$ is an adiabatic invariant and thus can be quantized according to the Ehrenfest principle, that classical adiabatic invariants may
correspond to observables with discrete spectrum. So the area of the horizon $A$ could be the proper candidate for the quantum mechanics of the black hole. 

Another candidate is  the gravitational coupling $K=1/4G$, where $G$ is the Newton constant.
It enters the Einstein-Hilbert gravitational action:
\begin{equation}
S_\text{grav}= 
 \frac{1}{4\pi }\int d^3x dt\sqrt{-g} K{\cal R}  \,,
\label{eq:gravity_action}
\end{equation}
where ${\cal R}$ is the scalar curvature.

In the modified gravity theories, such as the scalar-tensor and $f(R)$ theories 
(see e.g. \cite{Starobinsky1980} and the recent paper \cite{Barrow2020} with references therein), the effective Newton ``constant`` $G$ can be space-time dependent, and thus cannot be the fundamental constant. In the  superfluid $^3$He-A with Weyl fermionic quasiparticles, the coupling $K$ in the effective gravity and  the fine structure  ``constant`` $\alpha$ in the effective electrodynamics are  determined by physics on the microscopic level.\cite{Volovik2003} In these effective theories, one obtains $K\propto \Delta_0^2$ and $1/\alpha \propto \ln (\Delta_0/T)$, where $\Delta_0$ is the gap amplitude  and $T$ is the temperature of the liquid.  Here 
$\Delta_0$  and  $T$ play the role of the ultraviolet cut-off and of the infrared cut-off correspondingly.  In the inhomogeneous superfluid (inhomogeneous vacuum) both $K$ and $\alpha$ depend on coordinates.
It is not surprizing that in the relativistic quantum vacuum there is also the connection  between the gravitational coupling $K$ and 
the coupling $\alpha$  in quantum electrodynamics as suggested in Refs. \cite{Landau1955,Akama,AkamaTerazawa,Terazawa,TerazawaAkama,Terazawa1981,KlinkhamerVolovik2005}. 
 
 The Einstein-Hilbert action (\ref{eq:gravity_action}) contains the product of the scalar  Riemann curvature $R$  and the gravitational coupling constant $K$. This may suggest that these variables can be considered as the local thermodynamic variables, which are similar to temperature, pressure, chemical potential, number density, etc., in condensed matter physics.  Indeed, the Riemann curvature as the covariant quantity may serve as one of the thermodynamical characteristics of the macroscopic matter.\cite{Pronin1987}  If so,  the gravitational coupling constant $K$ in front of the scalar curvature  in the Einstein-Hilbert action also becomes the thermodynamic quantity, see Ref. \cite{KlinkhamerVolovik2008e}.
 
 Let us now go from the local  thermodynamic variables to the global ones, which characterize the global objects, such as the black hole.  Integration of the  Riemann curvature in Eq.(\ref{eq:gravity_action}) over the black hole Euclidean spacetime (the 4-volume integral)  gives rise to the global variable, which is canonically conjugate to $K$ in the black hole thermodynamics. It is the area $A$ of the black hole horizon.\cite{Volovik2020a} 
 
\subsection{Modified first law of black hole thermodynamics }
\label{FirstLaw}

Let us look  how the variable $K$ enters the thermodynamic laws for the black hole. In terms of this coupling $K$
the Hawking temperature of Schwarzschild black hole and its Bekenstein entropy are:
 \begin{equation}
  T_\text{BH}=\frac{K}{2\pi M}~~,~~S_\text{BH}
             = \frac{\pi M^2}{K}   \,.
\label{eq:HawkingT}
\end{equation}
 Then, using  the black hole area $A= \pi M^2/K^2$; the  gravitational coupling $K=1/4G$;  
the Hawking temperature $T_\text{BH}=M/2AK= K/2\pi M$; and the black hole entropy $S_\text{BH}=AK$, one obtains:
\begin{eqnarray}
dS_\text{BH}=d(AK)= \pi d (M^2/K)  =
\nonumber
\\
=- \pi \frac{M^2}{K^2}dK + 2\pi \frac{M}{K} dM \,.
\label{dS2}
\end{eqnarray} 

This suggests the following modification of the first law of black hole thermodynamics in case if $K$ is a global thermodynamic variable:
\begin{equation}
dS_\text{BH}= -AdK + \frac{dM}{T_\text{BH}}\,.
\label{FirstLawEq}
\end{equation}
This modification is similar to the modification in terms of the moduli fields.\cite{Gibbons1996} But in our case  the thermodynamic variable, which is  conjugate to the thermodynamic variable $K$, is the product of the black hole area and the black hole temperature, $AT_\text{BH}$. On the other hand in dynamics, $K$ and $A$ are canonically conjugate variables, see Sec.\ref{Canonical}. 

In general, the variable $K$ is local and depends on space coordinate, but in the same way as for the moduli fields,\cite{Gibbons1996} the black hole thermodynamics is determined by the asymptotic value of $K$ at spatial infinity. In Eq.(\ref{FirstLawEq}), $K\equiv K({\infty})$ is the global quantity, which characterizes the quantum vacuum in full equilibrium, i.e. far from the black hole. 

Also, the variable $K$ allows us to study the transition to the vacuum without gravity, i.e. to the vacuum where $K\rightarrow\infty$ and thus $G\rightarrow 0$, see Sec.\ref{Canonical}.

\subsection{Adiabatic change of $K$ and adiabatic invariant}
\label{adiabatic}

Let us change the coupling $K$ and the black hole mass $M$ adiabatically, i.e. at constant entropy of the black hole. Then the equation $dS_\text{BH}=0$ gives 
\begin{equation}
\frac{dM}{dK}= AT_\text{BH}=\frac{M}{2K} \,.
\label{ratio}
\end{equation}
This shows that $M^2/K={\rm const}$ is the adiabatic invariant for the spherical electrically neutral black hole. Thus according to  the Bekenstein conjecture \cite{Bekenstein1974}, it can be quantized in quantum mechanics:
\begin{equation}
\frac{M^2}{K}= a N \,.
\label{quantization}
\end{equation}
Here $N$ is integer, and $a$ is some fundamental dimensionless parameter of order unity.
If this conjecture is correct, one obtains the quantization of the entropy of Schwartzschild black hole:
\begin{equation}
S_\text{BH}(N)=\pi \frac{M^2}{K}= \pi a N \,.
\label{quantizationSchwarz}
\end{equation}

The Bekenstein idea on the role of adiabatic invariants in quantization of the black hole requires further consideration, see some approaches to that in Refs. \cite{Horowitz1996,Barvinsky2001,Barvinsky2002,Ansorg2012,Visser2012,Tharanath2013}.  In particular, the similarity between the energy levels of Schwarzschild black hole and the hydrogen atom has been suggested \cite{Bekenstein1997,Corda2015}.
We leave this problem for the future. This consideration should be supported by microscopic theory, see e.g. \cite{Carlip2014}. 
That is why the condensed matter analogs can be useful, since in the condensed matter systems the physics is known both on macro and micro levels.  

The consideration of the quantum nucleation of the vortex ring in moving superfluids --  the vortex instanton -- suggests that the corresponding entropy which determines the nucleation process, $\exp(-S_\text{ring})$,  is 
\begin{equation}
S_\text{ring}(N)=2\pi N \,.
\label{quantizationSchwarz}
\end{equation}
Here $N$ is the number of atoms involved in the process of the vortex instanton, see Sec. 26.4 and Eqs.(26.20)-(26.21) in Ref. \cite{Volovik2003}. This would correspond to $a=2$ in Eq.(\ref{quantization}).

The so-called $q$-theory can be also exploited, which allows us to consider dark energy, dark matter, black holes and the varying gravitational coupling in the frame of the same effective theory of the quantum vacuum
\cite{KlinkhamerVolovik2008e,Klinkhamer2017,Klinkhamer2019}.
 
\subsection{$A$ and $K$ as canonically conjugate variables and black-hole -- white-hole quantum tunneling}
\label{Canonical}

 The quantum mechanical  treatment of the black hole can be obtained,
if one finds the relevant canonically conjugate variables describing the black hole dynamics. Such approach has been suggested in Ref. \cite{Vagenas2011}, where the  canonically conjugate variables have been determined in the Euclidean time.  Since in Euclidean time the action for the black hole is equal to its entropy, $I_\text{E}=S_\text{BH}$,
in the theory with varying gravitational coupling $K$,   the proper  canonically conjugate variables are the gravitational coupling $K$ and the black hole area $A$. This allows us to consider the quantum mechanical tunneling from the black hole to the white hole -- the process  discussed in Refs. \cite{Barcelo2014,Barcelo2017,Rovelli2018,Rovelli2019,Rovelli2018b,Uzan2020b,Uzan2020c,Uzan2020d,Bodendorfer2019,Rovelli2021a} and in references therein -- in terms of the macroscopic quantum tunneling.  In  the semiclassical description of the quantum tunneling, the trajectory in the $(K,A)$ phase space is considered, which connects the black and white holes.

Note the difference from consideration in Sec. \ref{adiabatic}, where $K$ varies in the adiabatic regime, i.e. at fixed entropy $S_{\rm BH}$, while the area $A$, temperature $T_{\rm BH}$ and mass $M$ follow the variation of $K$.  In the dynamic regime, which is relevant for the description of quantum tunnelling, the parameter $K$ varies at fixed energy (fixed mass $M$ of the black hole), while the area $A$, the temperature $T_{\rm BH}$ and entropy $S_{\rm BH}$ follow the variation of $K$.

As in the case of the semiclassical consideration of the Hawking radiation in terms of the quantum tunneling \cite{Volovik1999,Wilczek2000,Volovik2009},
we shall use the Painleve-Gullstrand coordinate system \cite{Painleve,Gullstrand} with the metric:
 \begin{equation}
ds^2= - dt^2(1-{\bf v}^2) - 2dt\, d{\bf r}\cdot {\bf v} + d{\bf r}^2 \,.
\label{PGmetric}
\end{equation}
Here the vector $v_i({\bf r})=g_{0i}({\bf r})$ is the shift velocity -- the velocity of the free-falling observer, who crosses the horizon. 
In condensed matter the analog of this metric is the so-called acoustic metric,\cite{Unruh1981} emerging for quasiparticles in moving superfluids, where the shift velocity $v_i$ is played by superfluid velocity.  The analogs of the black hole and white hole horizons described by this metric can be also reproduced in the Dirac and Weyl topological semimetals,
where the horizon takes place on the boundary between different types of Dirac or Weyl materials.\cite{Volovik2016,Kuang2017,Zubkov2018,Wilczek2020,Sabsovich2021}

For the Schwartzschild black hole one has
\begin{equation}
{\bf v} ({\bf r})=\mp \hat{\bf r} \sqrt{\frac{R}{r}}=\mp \hat{\bf r} \sqrt{\frac{M}{2rK}}=\mp \hat{\bf r} \sqrt{\frac{2MG}{r}}\,,
\label{velocity}
\end{equation}
where $R$ is the radius of the horizon; the minus sign corresponds to the black hole and the plus sign describes the white hole.
In the theory with the variable gravitational coupling, the sign changes at the singularity $K=\infty$ (i.e. at $G=0$), when 
the black hole shrinks to a point and then expands as a white hole. The point $K=\infty$, where gravity disappears,  serves as the branch point, where the velocity of the freely falling observer changes sign. 

The vector ${\bf v}$ is normal to the surface of the horizon, and when it changes sign, this means that the horizon area $A$ changes sign: it crosses zero at $K=\infty$ and  becomes negative on the white-hole side of the process, $A \rightarrow -A$. This also could mean that due to connection between the area and entropy the white hole may have negative entropy. This is what we shall obtain in the following sections.

The quantum tunneling exponent is usually determined by the imaginary part of the action on the trajectory, which transforms the black hole to white hole. In terms of Euclidean action one has:
 \begin{equation}
p\propto \exp{\left(-I_{\text {BH} \rightarrow \text{WH}}\right)} \,\,,\,\, I_{\text{BH} \rightarrow \text{WH}} =  \int_C A(K')dK' \,.
\label{TunnelingExponentGen}
\end{equation}
Here the semiclassical trajectory $C$ is at $M=\rm{const}$, and thus $A(K')=\pm \pi M^2/(K')^2$. Along this trajectory the variable $K'$ changes from $K$ to the branch point at $K'=\infty$, and then from  $K'=\infty$ to $K'=K$ along the other  branch, where the area $ A(K')<0$.  
The integral gives the tunneling exponent of the transition to the white hole 
\begin{equation}
I_{\text{BH} \rightarrow \text{WH}}=  \,  2\pi M^2     \int_K^\infty \frac{dK'}{K'^2}= 2 \pi \frac{M^2}{K}  \,.
\label{TunnelingExponent}
\end{equation}
The tunneling exponent in Eq.(\ref{TunnelingExponent}) can be expressed in terms of the black hole entropy in Eq (\ref{eq:HawkingT}), and for the probability of transition one obtains:
\begin{equation}
p\propto \exp{\left(-2\pi M^2/K\right)}=\exp{\left(-2S_\text{BH}\right)}
\,.
\label{tunneling}
\end{equation}
Note that it is twice the black hole entropy, which enters Eq.(\ref{tunneling}).

\subsection{Black hole to white hole transition as series of the Hawking radiation cotunneling}
\label{cotunneling1}

Let us show that the result (\ref{tunneling}) with twice the black hole entropy is supported by consideration of the conventional Hawking radiation of particles from the black hole. 
The tunneling exponent of radiation can be expressed in terms of the change in the entropy of the black hole after radiation,  $p\propto e^{\Delta S_{\text{BH}}}$, see Refs. \cite{Kraus1997,Parentani1997,Berezin1999,Wilczek2000}. This demonstrates that the Hawking radiation process has the thermodynamic meaning as thermodynamic fluctuation.\cite{Landau_Lifshitz}  

We consider the process of the co-tunneling, in which the particle escapes the black hole by quantum tunneling, and then this particle tunnels to the white hole through the white hole horizon (this is the analog of the electron tunneling via an  intermediate virtual state in electronic systems \cite{Feigelman2005,Glazman2005}). This process takes place at the fixed total mass $M$. The tunneling exponent for this process  to occur is $e^{(\Delta S_{\text{BH}}+\Delta S_{\text{WH}})}$. 
Summation of all the processes of the tunneling of matter from the black hole to the formed white hole gives finally Eq.(\ref{tunneling}):
\begin{equation}
p\propto e^{\sum(\Delta S_{\text{BH}}+\Delta S_{\text{WH}})}=e^{2\sum\Delta S_{\text{BH}}}=\exp{\left(-2S_\text{BH}\right)}
\,.
\label{tunneling2}
\end{equation}
Here we took into account the (anti)symmetry in the dynamics of black and white holes in the process of quantum tunneling,
$\sum\Delta S_{\text{BH}}=\sum \Delta S_{\text{WH}}$.

\subsection{Emission of small black holes vs Hawking radiation}
\label{EmissionSmall}

The principle that the macroscopic quantum  tunneling can be considered as thermodynamic fluctuation can be also applied to the process of the creation of  pairs black holes,\cite{HawkingHorovitz1995} to the process of splitting of the black hole into two or several smaller black holes with the same total mass, see e.g. Ref.\cite{HyeyounChung2011}, and to other processes with macroscopic objects. For example, the probability of the decay  of the black hole with mass $M$ into two black holes with $M_1+M_2=M$ is:
\begin{eqnarray}
p(M \rightarrow M_1+M_2) \propto \,
\\
e^{S_{\text{BH}}(M_1)+S_{\text{BH}}(M_2)-S_{\text{BH}}(M_1+M_2)} =e^{-8\pi G M_1M_2}.
\label{tunnelingM1M2}
\end{eqnarray}

In the particular limit case $m=M_1 \ll M_2\approx M$, this channel of the black hole decay describes the emission of the small black hole with mass $m$ by the large black hole with mass $M$:
 \begin{equation}
p(m,M)\propto \exp{\left(-\frac{m}{T_{\rm BH}(M)}\right)}  \,\,,\,\, m \ll M
\,.
\label{tunnelingMm}
\end{equation}
This shows that the macroscopic tunneling process of emission of a small black hole  by a large black hole is governed by the same Hawking temperature 
$T_{\rm BH}(M)=1/(8\pi GM)$ as the Hawking radiation of a particle, which tunnels across the horizon. 

However, there is the difference.
In Eq.(\ref{tunnelingMm}) the quadratic term $m^2$ is neglected. In general case, one obtains from Eq.(\ref{tunnelingM1M2}):
 \begin{equation}
p(m,M-m) \propto \exp{\left(-8\pi Gm(M-m)\right)}
\,.
\label{tunnelingMmquadratic}
\end{equation}
This equation demonstrates the effect of back reaction -- the correction to the Hawking radiation caused by the reduction of the black hole mass after the radiation of a small black hole.
The similar correction due to the back reaction in the process of radiation of particles was obtained by Parikh and Wilczek \cite{Wilczek2000} (see also Ref.\cite{Kraus1997}):
 \begin{equation}
p(\omega, M-\omega)\propto \exp{\left(-8\pi G\omega\left(M-\frac{\omega}{2}\right)\right)}
\,,
\label{tunnelingMomega}
\end{equation}
where $\omega$ is the energy of the emitted particle.
As distinct from Eq.(\ref{tunnelingMmquadratic}) for emission of a small black hole, the Eq.(\ref{tunnelingMomega}) contains the factor $1/2$.
The reason for such difference is that in case of the emission of a small black hole, the probability of emission contains the extra term compared to the emission of particles -- the entropy of the emitted black hole, $S(m)=4\pi G m^2$:
 \begin{equation}
p(m,M-m)\propto \exp{\left(-8\pi Gm\left(M-\frac{m}{2}\right)+ 4\pi G m^2\right)}
\,.
\label{tunnelingMm2}
\end{equation}
As a result the Eq.(\ref{tunnelingMmquadratic}) is restored. This also coincides with the result for the radiation of the self-gravitating shell.\cite{KrausWilczek1995}

The back reaction is also important for the process of the vortex creation by macroscopic quantum tunnelling. The calculations by Sonin\cite{Sonin1973} demonstrate the logarithmic deviations  from the simple equation (\ref{quantizationSchwarz}), which does not take into account the back reaction of the  flow of the liquid.

Eq.(\ref{tunnelingM1M2}) can be extended to emission of several black holes. For example, the probability of emission of $N$ black holes with masses $m=M/N$ is:
\begin{eqnarray}
p(M \rightarrow mN) \propto \exp{\left(-4\pi G M^2\left(1 -\frac{1}{N}\right)\right)}\,.
\label{tunnelingMN}
\end{eqnarray}
For large $N$ this corresponds to the probability of the destruction of the black hole by explosion within the particular channel:
\begin{eqnarray}
p_{N\rightarrow \infty} \propto \exp(-S_{\rm BH})\,.
\label{tunnelingMNinf}
\end{eqnarray}
There are many channels of the destruction of the black hole in quantum tunnelling. The black hole entropy  $S_{\rm BH}$ can be estimated using the probability  of the black hole explosion in a single quantum event. Eq.(\ref{tunnelingMNinf}) suggests that the entropy of the  black hole can be determined by the number of channels of its destruction.

Emission of the smaller black holes is typically discussed in relations to the near-extremal black holes, see e.g. Ref.\cite{McInnes2021}
and references therein.

\subsection{Negative entropy of white hole}
\label{WHentropy}

The connection between the probability of the transition and the thermodynamic fluctuations\cite{Landau_Lifshitz} can be applied  to the transition between the black and white holes.  The total change of the entropy in this  process is
$\Delta S=S_\text{WH}-S_\text{BH}$. According to Eq.(\ref{tunneling2}) this change is  $\Delta S=-2S_\text{BH}$. As a result, one obtains that the entropy of the white hole is equal with the opposite sign to the entropy of the black hole with the same mass 
\begin{equation}
S_\text{WH}(M)=-S_\text{BH}(M)
\,.
\label{fluctuation}
\end{equation}
This means that the white hole, which is obtained by the quantum tunneling from the black hole and thus has the same mass $M$ as the black hole, has the negative temperature 
 $T_\text{WH}=- T_\text{BH}$ and the negative area $A_\text{WH}=- A_\text{BH}$, which together produce the negative entropy $S_\text{WH}=- S_\text{BH}$.
 
 In the black hole physics the negative temperature has been discussed for the inner horizon of the Kerr and charged black holes, see Ref.\cite{Gibbons2018} and references therein. The entropy of the inner horizon has been considered as positive.
 However, the arguments in Ref.\cite{Gibbons2018} do not exclude the possibility of the opposite situation, when the temperature of the inner horizon is positive, while the entropy is negative: the equation $T_+S_++T_-S_-=0$ remains valid.

 The discussed transition from the black hole to the white hole with the same mass $M$ is not the thermodynamic transition. It is the quantum process of tunneling between the two quantum states. It is one of many routes of the black hole evaporation, including formation of small white hole on the late stage of the decay \cite{Barcelo2014,Barcelo2017,Rovelli2018,Rovelli2019,Rovelli2018b,Uzan2020b,Uzan2020c,Uzan2020d,Bodendorfer2019,Rovelli2021a}.
The uniqueness of this particular route, the hidden information and (anti)symmetry between the black and white holes are combined to produce the negative entropy of the white hole, see also discussion in Sec.\ref{TunnelingNegEntropy}.

  \subsection{Negative temperature}

The negative temperature is a well defined quantity in condensed matter, see review Ref. \cite{Baldovin2021}. It typically takes place in the subsystem of nuclear spins, where the energy is restricted from above. Different  thermodynamic phase transitions occurring at $T<0$ have been experimentally observed in magnetic systems, see e.g. Ref. \cite{Hakonen1992}. The negative energy states are unstable both before and after the magnetic phase transition. This is because the state with $T<0$ is hotter than the state with $T>0$, since the heat flows from the negative- to the positive-temperature system. This means that if the black hole and white hole are in some contact,  the heat will escape from the white hole and will be absorbed by the black hole.

In our case, the transition from positive to negative $T$ occurs on the path $A(K)dK$ in Eq.(\ref{TunnelingExponentGen}) via the point $K=\infty$, i.e. via the infinite temperature $T(K=\infty)=\infty$. The transition via the infinite temperature is the analytic route to the thermodynamically unstable states with negative $T$, see e.g. Refs. \cite{Volovik2019,Volovik2019b}, where the transition to anti-spacetime\cite{Sakharov1980,Christodoulou2012,Boyle2018,Boyle2018b,Boyle2021} has been considered. In spin systems the $T<0$ state is typically obtained by reversing the magnetic field, which looks like crossing $T=0$.

 At first glance the state with the negative temperature is unstable. However, it is unstable only in case if there is the normal environment -- the thermal bath with positive temperature.  If there is no external environment, i.e. this "false" vacuum occupies the whole universe, this isolated vacuum becomes stable in spite of its negative temperature.

In the relativistic physics, the energy is unbounded. The negative and positive energy branches of fermionic states are  symmetric with respect to zero energy. Due to this symmetry the isolated "false" relativistic vacuum, in which all the positive energy states are occupied and the negative energy states are empty, has the same physics as the conventional Dirac vacuum in the "normal" Universe. 
Though the matter (excitations)  in this "false" Universe has negative energy, and the thermodynamic states are characterized by the negative temperature, the inhabitants of the "false" Universe think that they live in the "normal" Universe with positive energies for matter and positive temperature. It is only with respect to our Universe their temperature and energies are negative. But with respect to their Universe it is our Universe, which is false and which is described by negative temperature. 
So, in principle, we cannot resolve, whether our Universe is "false" or "true", see also discussion in Ref. \cite{Volovik2021b}.

\subsection{Discussion II}

We considered the dynamics and thermodynamics of the black hole in case of the varying gravitational coupling $K$. The gravitational coupling $K$ serves as the thermodynamic variable, which is thermodynamic conjugate to $A_\text{BH}T_\text{BH}$, where 
$A_\text{BH}$ and $T_\text{BH}$ are correspondingly the area of the black hole horizon and Hawking temperature.
The corresponding first law of the black hole is modified, see Eq.(\ref{FirstLawEq}). The corresponding adiabatic invariant is  the entropy $S_\text{BH}=KA_\text{BH}$, and it is this invariant which should be quantized, if the Bekenstein conjecture is correct.
This is in agreement with  observation of Ted Jacobson \cite{Jacobson1994}, that it is the entropy that does not change under renormalization of $K$, rather than the area.  This suggests the alternative quantization scheme for the black hole.  While $K$ and $A_\text{BH}$ are dimensionful  and cannot be quantized, the entropy is dimensionless and thus can be quantized in terms of some  fundamental numbers.   

On the quantum level the dynamically conjugate variables of the black hole physics are $K$ and $A_\text{BH}$. This allows us to consider
the transition from the black hole to the white hole as quantum tunneling in the semiclassical approximation, which is valid when the action is large. The classical trajectory of the black hole crosses the branch point at $K=\infty$, and then continues along the other branch, where the area $ A(K)<0$, which corresponds to the white hole. The obtained tunneling exponent $\exp(-2S_\text{BH})$ demonstrates that the transition can be considered as thermodynamic fluctuation, if the entropy of the white hole (with the same mass $M$  as the black hole) is negative, $S_\text{WH}=- S_\text{BH}$.

\section{Black hole to white hole quantum tunneling via coordinate transformation}
\label{coordinate}

The negative entropy is a rather unusual phenomenon. That is why to check that such conjecture may have sense, one should try different methods of calculations of the quantum tunneling from the black hole and to the white hole. In previous Section the tunneling exponent was calculated using the path in the complex plane of the varying gravitational coupling (the inverse Newton "constant"). 
Here we consider the intermediate state between the black hole and white hole, which is represented by the fully static metric, and calculate the transition probability from black hole to static hole and then from static hole to the white hole.\cite{Volovik2021f}  We shall use the path on the complex coordinate plane, which connects the static coordinates with the stationary but not static coordinates of the black or white holes. As a result the equation $S_\text{WH}(M)=- S_\text{BH}(M)$ will be confirmed, while for the intermediate symmetric static hole the zero entropy will be obtained.

Let us start with Schwarzschild black hole.

\subsection{Black and white holes as degenerate states of violated time reversal symmetry}

The black and white holes have stationary but not static metric, which in the Painleve-Gullstrand\cite{Painleve,Gullstrand}  (PG) coordinate system have the form in Eq.(\ref{PGmetric}).
For the Schwartzschild black hole one has
\begin{equation}
{\bf v} ({\bf r})=\mp \hat{\bf r} \sqrt{\frac{R}{r}}\,,
\label{velocity2}
\end{equation}
where $R=2M$, and again the minus sign corresponds to the black hole and the plus sign describes the white hole. In both states the time reversal symmetry is violated, since $T{\bf v}= - {\bf v}$, and thus under the time reversal operation the black hole transforms to the white hole.

So, the black and white holes represent two degenerate states of the object with the same mass $M$, which corresponds to the spontaneously broken time reversal symmetry.
This suggests that there can be also the  intermediate symmetric state, in which the time reversal symmetry is not violated, i.e. the fully static configuration with zero shift velocity, the grey black hole in the language of Refs.\cite{Hajicek2001,HajicekKiefer2001,Kiefer2019}.

\subsection{Fully static hole as symmetric intermediate state}

The role of such static state can be played by the configuration described by the fully static metric:
\begin{equation}
ds^2=- \left(  1- \frac{R}{r} \right)dt^2 + \frac{dr^2}{\left(  1- \frac{R}{r} \right)} + r^2 d\Omega^2
\,.
\label{StaticMetric}
\end{equation}
This metric has coordinate singularity at the horizon, but is invariant under time reversal.
Such static hole has the same mass as the black hole state. We consider this state as the intermediate state in the quantum tunneling from the black hole (BH) to white hole (WH). The probability of such transition  via the static state has probability:
\begin{equation}
P_{\text{BH}\rightarrow \text{WH}} = P_{\text{BH}\rightarrow \text{static}}P_{\text{static}\rightarrow \text{WH}}\,,
\label{transition1}
\end{equation}
with $P_{\text{BH}\rightarrow \text{static}}=P_{\text{static}\rightarrow \text{WH}}$ as follows from symmetry between black and white holes. 

\subsection{Tunneling from black hole to static hole and zero entropy of static hole}

Let us calculate $P_{\text{BH}\rightarrow \text{static}}$ using again the semiclassical quantum tunneling method, but in  the approach  related to coordinate transformations between the different states of the hole object. 

The transformation from static hole with metric (\ref{StaticMetric}) to the black hole with PG metric (\ref{PGmetric}) is obtained by the singular coordinate transformation:
\begin{equation}
dt \rightarrow d\tilde t +dr \frac{v}{1-v^2} \,.
\label{CoordinateTransformation}
\end{equation}
Such singularity in the coordinate transformation drastically changes the action $\int M dt$, which demonstrates that  the static hole and the black hole are absolutely different objects. However, in quantum mechanics these two states can be connected by path on the complex coordinate plane, which can be considered as the path along the trajectory describing the macroscopic quantum tunneling between the two hole objects. 

The probability of the quantum tunneling from the black hole to its static partner with the same mass $M$ is given by the imaginary part in the action on this path:
\begin{eqnarray}
 P_{\text{BH}\rightarrow \text{static}} = \exp{ \left(-2{\rm Im} \int M dt \right)} = 
 \label{ImaginaryAction1}
  \\
 = \exp{ \left(-2{\rm Im} \int \left(M d\tilde t + dr \frac{Mv}{1-v^2} \right) \right)} =
\nonumber
 \\
 = \exp{ \left(-2{\rm Im} \int dr \frac{Mv}{1-v^2}  \right)} =
 \label{ImaginaryAction2}
 \\
 =  \exp{ \left(-2\pi M R\right)} =  \exp{ \left(-4\pi M^2G\right)} 
 =  \exp{ \left(-S_\text{BH} \right)} \,.
\label{ImaginaryAction}
\end{eqnarray}

Note that in the above derivation the conventional Hawking radiation of particles from the black hole is not taken into account. Here only one channel of the decay of the black hole is considered: the macroscopic tunneling to the static hole (and then to the white hole). The combined many-channel decay is for further consideration in future. In the absence of Hawking radiation, the mass $M$ of the hole object is conserved, and thus the action is simply $\int M dt$, which acquires the imaginary part from the singularity in the coordinate transformation. 

From Eq.(\ref{ImaginaryAction}) it follows that the probability of the macroscopic quantum tunneling from black hole to the static state is fully determined by the black hole entropy $S_\text{BH}$. On the other hand the quantum tunneling transition can be considered  as quantum fluctuation and thus it is determined by the difference in the entropy of the initial and final states:\cite{Landau_Lifshitz}
\begin{equation}
P_{\text{BH}\rightarrow \text{static}} =\exp{ \left(S_\text{static} -S_\text{BH} \right)} \,.
\label{transition5}
\end{equation}
Then from Eqs.(\ref{ImaginaryAction}) and (\ref{transition5})  it follows that the entropy of the static  hole is zero, $S_\text{static}=0$. This is rather natural, because here the Schwarzschild  static solution of Einstein equations described by metric (\ref{StaticMetric}) is considered not as the black hole, but as the intermediate grey state between the black and white holes, which has two independent static patches. The intermediate state has no Hawking radiation, its horizon represents  the surface of infinite red shift. This static hole has no absorption or emission of particles, instead there is a reflection of particles from such horizon on both sides of the coordinate singularity. This state should have  zero temperature.

Here we note that in general relativity not all coordinate transformations are allowed. There are some singular transformations, which change the state of the system, i.e. they transfer the initial state  to the final state, which is physically different from the initial state. The transformation from the PG coordinates for black hole to the PG coordinates for white hole belongs to this class. BH and WH are physically different systems, with different entropies and temperatures. The other situations, when two frames are not thermodynamically equivalent, have been discussed in Refs.\cite{Odintsov2020,Odintsov2017}.

The same concerns the singular transformation from the PG coordinates for black hole to the fully static Schwarzschild coordinates. This transformation produces the intermediate static state, which is physically not the BH and not the WH. This does not contradict to the consideration of the black hole in Schwarzschild coordinates. When the real black hole is considered using Schwarzschild coordinates, the singularity is usually avoided by some tricks, which correspond to the black hole  behaviour at the horizon (or  to the white hole behaviour at the horizon if the white hole is considered). But here we consider the truly static state as the real solution of Einstein equations, which serves as the intermediate state between the black hole and the white hole. The particles propagating at this static spacetime are also considered without ascribing the black or white hole features to the horizon. 
This means that all the three states (black, white and intermediate static), which cannot be obtained from each other by the regular coordinate transformations, are physically different.

\subsection{Tunneling from black hole to white hole and negative entropy of white hole}
\label{TunnelingNegEntropy}

Let us now consider  the quantum tunneling from black hole to white hole. From Eqs. (\ref{transition1})  and (\ref{ImaginaryAction}) it follows that the probability of the tunneling is:
\begin{equation}
P_{\text{BH}\rightarrow \text{WH}} = P^2_{\text{BH}\rightarrow \text{static}} =  \exp{ \left(-8\pi M^2G\right)} =\exp{ \left(-2S_\text{BH} \right)} .
\label{transition3}
\end{equation}
This reproduces Eq.(\ref{tunneling}) and thus confirms the negative entropy of the white hole.
The same result (\ref{transition3}) can be obtained using the direct transformation from the black hole to the white hole:
\begin{equation}
dt \rightarrow dt +2dr \frac{v}{1-v^2} \,.
\label{CoordinateTransformationBHWH}
\end{equation}

In principle, the equation of the type $\exp{ \left(-\gamma M^2G\right)}$ for the probability of the black hole to white hole tunneling is expected on dimensional grounds, see e.g. Ref.\cite{Rovelli2018}. However, the value of the parameter $\gamma=8\pi$ in Eq.(\ref{transition3}) leads to the unexpected result -- the negative entropy of the white hole in Eq.(\ref{fluctuation}). The white hole entropy is with minus sign  the entropy of the black hole  with the same mass. 

The black hole states with negative entropy have been considered in Ref.\cite{Odintsov2002}, where it has been suggested that appearance of negative entropy may indicate a new type instability, see also Ref.\cite{Odintsov2020}.
Such super-low entropy of white hole can be also seen as an example of a memory effect discussed in Ref.\cite{Rovelli2020}, i.e.
the entropy of the white hole is negative, since this state remembers that it is formed from the black hole by the quantum tunneling.

Let us consider the memory effect using the following unitary cycle.\cite{Hajicek2001} We start with the thin shell of matter with zero entropy. In the first step the shell collapses to the black hole, in the second step the black hole tunnels quantum mechanically to the white hole, and finally in the third step the shell is released from the white hole. If one assumes that the information on the shell is not lost inside the black hole, the restored shell will have again zero entropy. But this shell is formed from the white hole, which suggests that the white hole state has negative entropy. In this cycle the entropy increases from zero to $S_{\rm BH}$ in the first step, then it decreases from $S_{\rm BH}$ to $-S_{\rm BH}$ in the quantum tunneling, and finally increases from $-S_{\rm BH}$ to zero in the third step. This demonstrates, that the entropy of the black and white holes is closely related to the information puzzle.\cite{Hawking2015} 

Since the white hole is obtained from black hole by the quantum process, one can consider the entropy of this state as the conditional entropy in the quantum entanglement of these two states. Such conditional entropy of the white hole is with minus sign of the entropy of the black hole.

\subsection{Discussion III}

The method of the calculations of the macroscopic quantum tunneling exponent using  the singular coordinate transformations produced the same results, which were obtained using the variable gravitational coupling $K$. That is why we can move further and consider the application of the method to the other types of the black holes.

\section{Black holes with two horizons}
\label{TwoHorizons}

\subsection{Introduction}

For the simplest (Schwarzschild)  black hole the Bekenstein-Hawking entropy is proportional to the area of the event horizon, $S=A/4G$.
However, for the black hole with two or more horizons, such as the rotating and charged black holes and the black holes in the de Sitter space-time, the thermodynamics is not well established, since the role of the inner horizon remains unclear. 
The inner  horizon may experience instabilities,\cite{Penrose1968,McNamara1978,Matzner1979,Starobinsky1979,Poisson1990,Hod1998,Burko1998,Chesler2020}
which may or may not produce singularity at the horizon, or even prevent the formation of the inner horizon.
If the inner horizon is present it may contribute to the entropy of the black hole. However,
the total entropy of the system is not necessarily the sum of the entropies of the horizons, since the correlations between the horizons are possible, see e.g. Refs.\cite{YunHe2018,YuboMa2021,Odintsov2021}.

Here we consider the spherically symmetric black hole with two horizons -- the charged Reissner-Nordstr\"om black hole, and ignore the possible instabilities of the inner horizon. We  calculate the  entropy and the corresponding temperature of the thermal Hawking radiation using three different approaches. All of them give the same results, which demonstrate the correlations between the horizons.

In Sec.\ref{HawkingRadiation} the radiation from the RN black hole is considered using the semiclassical 
tunneling approach.\cite{Wilczek2000,Srinivasan1999,Volovik1999,Akhmedov2006,Vanzo2011,Jannes2011}
When both inner and outer horizons are taken into account, one obtains that the temperature of radiation does not coincide with the conventional Hawking temperature related to the outer horizon. The modified Hawking temperature does not depend on the charge $Q$, and thus is the same as in case of the Schwarzschild black hole with the same mass $M$, i.e. $\tilde T_{\rm H} = \frac{1}{8\pi M}$.

In Sec.\ref{Interpretation} it is shown that this result can be written as the product of the thermal factors, determined by the Hawking temperatures of two horizons: $\exp{\left(-\frac{E}{\tilde T_{\rm H} }\right)} =
\exp{\left(-\frac{E}{T_+}\right)}  \times \exp{\left(-\frac{E}{T_-}\right)}$. This demonstrates that the Hawking radiation in the presence of two horizons is the correlated process, which is similar to the phenomenon of cotunneling in the electronic systems, when an electron experiences the coherent sequence of tunneling events:  from an initial to the virtual intermediate states and then to the final state.\cite{Feigelman2005,Glazman2005} In our case the virtual intermediate state is in the region between the two horizons.

In Sec.\ref{Singular} the entropy is obtained in the approach based on the singular coordinate transformations. The result for the total entropy,  $\tilde S_{\rm RN}= 4\pi M^2$, also does not depend on charge $Q$ and is the same as for neutral black hole with the same mass. This is in agreement with the modified Hawking temperature due to thermodynamic relation $d\tilde S_{\rm RN} =  d(4\pi M^2)=\frac{dM}{\tilde T_{\rm H}}$.

The independence of the entropy on charge $Q$ suggests that the states with different $Q$ can be obtained by the adiabatic transformations. In Sec.\ref{Adiabatic} we consider such adiabatic transformation, which does not contradict to the conservation of the charge $Q$. It corresponds to adiabatic change of the fine structure constant $\alpha$ to its zero value.
When $\alpha$ slowly decreases to zero the two horizons move slowly  with conservation of the charge number $Q$ and energy $M$. 
Finally the inner horizon disappears and the black hole at $\alpha=0$ becomes neutral. In such slow process the entropy does not change, and is the same as the entropy of the neutral black hole. 

\subsection{Reissner-Nordstr\"om  static hole}

Let us start with the fully static metric of Reissner-Nordstr\"om black hole:
  \begin{equation}
ds^2= - dt^2 a(r)   + dr^2\frac{1}{a(r)} +r^2 d\Omega^2.
\label{staticRN1}
\end{equation}
where
 \begin{equation}
a(r) =\frac{(r-r_-)(r-r_+)}{r^2} \,.
\label{staticRN}
\end{equation}
Here the event horizon is at $r=r_+$  and the inner horizon is at $r=r_-$.
The positions of horizons  are expressed in terms of the mass $M$ and charge $Q$ of the black hole  ($G=c=1$):
 \begin{equation}
r_+r_-= Q^2\,\,, \,\, r_+ + r_-= 2M\,.
\label{r+r-}
\end{equation}

The conventional entropy of RN black hole is typically related to the area of the outer horizon:
 \begin{equation}
S_{\rm RN}(r_+)= \pi r_+^2= \pi \left( M + \sqrt{M^2 - Q^2}\right)^2
\label{entropyRN}
\end{equation}
However, in Ref.\cite{Volovik2021d} the different result was obtained, i.e. the entropy does not obey the area law and does not depend on $Q$: 
 \begin{equation}
\tilde S_{\rm RN}=\pi (r_+ + r_-)^2= 4\pi M^2 \,.
\label{entropyRNtilde}
\end{equation}
Here we explain this behaviour of the entropy using several different approaches.

\subsection{Temperature of Hawking radiation from black hole with two horizons is universal}
\label{HawkingRadiation}

The Eq.(\ref{entropyRNtilde}) can be obtained from the temperature of the thermal Hawking radiation, which can be found 
using the method of the semiclassical quantum tunneling.\cite{Wilczek2000,Srinivasan1999,Volovik1999,Akhmedov2006,Vanzo2011}
 Now we have two horizons, and  thus the Hawking radiation temperature can be modified.   In the semiclassical method the process of Hawking radiation is typically studied using the Painleve-Gullstrand (PG) metric,\cite{Painleve,Gullstrand} which has no singularity at the horizon. However, for the RN black hole the PG coordinates are not well defined due to the existence of the inner horizon,\cite{Faraoni2020} and the extension of the PG coordinates is required.\cite{Hennigar2022} The proper extension of the PG  metric is obtained by the following coordinate transformation:\cite{Volovik2003b}
 \begin{equation}
dt \rightarrow dt \pm fdr  \,\,,\, f=\frac{\sqrt{2Mr}}{\sqrt{r^2 + Q^2}}\frac{r^2}{(r-r_-)(r-r_+)}\,.
\label{CoordinateTrnaformationRN2}
\end{equation}
The corresponding black hole and white hole states have the following metric:
 \begin{equation}
ds^2= - dt^2\left( 1 +\frac{Q^2}{r^2}\right)+ \frac{1}{1 +\frac{Q^2}{r^2}} (dr\pm v dt)^2+r^2 d\Omega^2\,.
\label{PG_RN2}
\end{equation}
Here the shift velocity is determined by equation:
 \begin{equation}
v^2 = \frac{2M}{r} \left( 1 +\frac{Q^2}{r^2}\right)\,,
\label{velocity_RN2}
\end{equation}
and $+$ and $-$  signs correspond to black and white holes.
In this extension of the PG coordinates, the shift velocity is real for all $r$. This is distinct from the conventional  (non-modified) PG coordinates,\cite{Hamilton2008,Zubkov2019} where the shift velocity becomes imaginary for $r<r_0= r_+r_-/(r_+ +r_-)$.

 The tunneling trajectory for a massless particle 
can be found from the energy spectrum of a particle. The latter is determined by  the contravariant metric $g^{\mu\nu}$, which is inverse to the PG metric $g_{\mu\nu}$ in Eq.(\ref{PG_RN2}):
 \begin{equation}
g^{\mu\nu}p_\mu p_\nu=0 \,\, \rightarrow  E = p_r v(r) \pm p_r \left(1 + \frac {Q^2}{r^2}\right)\,.
\label{InverseMetric}
\end{equation}
Here $p_r$ is the radial momentum, and $E=p_0$. 

The probability of the tunneling process of the Hawking radiation is given by the exponent of  imaginary part of the action along the tunneling trajectory of particle, ${\rm Im} \int p_r(r,E) dr$,
where the trajectory $p_r(r,E)$ is 
 \begin{equation}
p_r(r,E) =\frac{E}{ v(r) -\left(1 + \frac {Q^2}{r^2}\right)}\,.
\label{pr}
\end{equation}
In the system with two horizons, the imaginary part of the action is produced by both poles. The contribution of two poles gives the following probability of the Hawking radiation:
 \begin{eqnarray}
P= \exp\left( - 4\pi E \left(  \frac{r_+^2}{r_+ -r_-} - \frac{r_-^2}{r_+ -r_-} \right)  \right) =
\label{P1}
\\
=  \exp\left( - 4\pi E (r_+ +r_-) \right).
\label{P2}
\end{eqnarray}
This corresponds to the thermal radiation characterized by the following Hawking temperature:
 \begin{equation}
\tilde T_{\rm H} = \frac{1}{4\pi (r_+ +r_-)}= \frac{1}{8\pi M} \,.
\label{HawkingTtilde1}
\end{equation}
This differs from the traditional Hawking temperature  determined by the outer horizon, $T_{\rm H} = \frac{1}{4\pi r_+}$.
The modification is caused by the contribution from the inner horizon.

The modified Hawking temperature in Eq.(\ref{HawkingTtilde1}) is universal. It is fully determined by the mass $M$ of the RN black hole, and thus does not depend on the charge $Q$ of the RN black hole. This result was also suggested in Ref. \cite{Makela2001} based on the consideration of the discrete spectrum.

 \subsection{Universality of entropy}
 \label{UniversalEntropy}

The independence of the modified Hawking temperature $\tilde T_{\rm H}$ on charge $Q$ supports the Eq.(\ref{entropyRNtilde}), which suggests that  the entropy $\tilde S_{\rm RN}$ of the charged black hole is universal: it also does not depend on $Q$. 
The temperature  $\tilde T_{\rm H}$  and entropy $\tilde S_{\rm RN}$  satisfy the thermodynamic equation:
 \begin{equation}
d\tilde S_{\rm RN} =  d(4\pi M^2)=\frac{dM}{\tilde T_{\rm H}} \,.
\label{ThermodynamicsTtilde}
\end{equation}

It is important that the Hawking temperature $\tilde T_{\rm H}$ remains constant, when the black hole approaches the extremal limit. That is why
contrary to the proposal in Ref. \cite{Hod2021}, the radiation quanta emitted by the near-extremal  RN black hole cannot transform it into the naked singularity, and thus the Penrose cosmic censorship conjecture is not violated.\cite{Penrose1969}

Anyway, it would be interesting to consider the limit of extremal black hole.
 In the extremal limit $r_+ - r_-  \rightarrow 0$, the black hole entropy may finally drop from $4\pi M^2$ to zero. The possible change of the regime in the extreme limit was considered in Ref.\cite{Volovik2003b}.

 \subsection{Interpretation in terms of Hawking temperatures at two horizons}
 \label{Interpretation}
 
Equation (\ref{P1}) can be interpreted in terms of the conventional Hawking temperatures at two horizons: it can be written as  the product of two thermal factors:
  \begin{equation}
P=P_+P_-=\exp{\left(-\frac{E}{T_+}\right)}  \times \exp{\left(-\frac{E}{T_-}\right)}  \,.
\label{cotunneling}
\end{equation}
Here $T_+$ and $T_-$ are the conventional Hawking temperatures at two horizons, which are determined by the gravity at the horizons:\cite{ZhaiLiu2010}
\begin{eqnarray}
T_+= \frac{1}{4\pi} a'(r_+)= \frac{1}{4\pi} \frac{r_+ -r_-}{r_+^2} \,,
\label{T+}
\\
T_-= \frac{1}{4\pi} a'(r_-)= -\frac{1}{4\pi}  \frac{r_+ - r_-}{r_-^2} \,.
\label{T-}
\end{eqnarray}
The  temperature $T_+$ determines the rate of the tunneling from the region $r_-<r<r_+$ to $r>r_+$, while the negative  temperature $T_-$ determines the occupation number of these particles near the inner horizon.

Altogether the correlated effect of two horizons leads to the modified Hawking temperature in Eq.(\ref{HawkingTtilde1}):
  \begin{eqnarray}
P=P_+P_- = \exp{\left(-\frac{E}{\tilde T_{\rm H}} \right)} \,,
\label{cotunneling}
\\
\tilde T_{\rm H}= \left( \frac{1}{T_+} -   \frac{1}{|T_-|} \right)^{-1} \,.
\label{EffT}
\end{eqnarray}
The correlated process in Eq(\ref{cotunneling}) is similar to the phenomenon of cotunneling in the electronic systems -- the coherent process of electron tunneling  from an initial to final state via virtual intermediate state,\cite{Feigelman2005,Glazman2005} which we discussed in Sec. \ref{cotunneling1} .
In a given case the virtual intermediate state is in the region between the two horizons.

Eq.(\ref{EffT}) has been also suggested for the Schwarzschild-de Sitter black hole,\cite{LiMaMa2017,Nakarachinda2017}  with $T_+$ and $T_-$ being the  temperatures of the cosmological horizon and the black hole correspondingly.

\subsection{Entropy in approach based on the singular coordinate transformations}
\label{Singular}

The same expressions for the entropy and the Hawking temperature of the RN black hole can be obtained using the approach in Sec.\ref{coordinate}. Now we discuss the macroscopic quantum tunneling  from the non-static PG RN black hole in Eq.(\ref{PG_RN2}) to the static state in Eq.(\ref{staticRN1}). The latter is viewed as intermediate state between the black hole and white holes states in Eq.(\ref{PG_RN2}). Assuming that the intermediate fully static state  has the natural zero entropy, one obtains the entropy of the black and white holes.  

The probability of the quantum tunneling from the RN black hole to its static RN partner in Eq.(\ref{staticRN1}) with the same mass $M$ and charge $Q$ is determined by singularities in the coordinate transformation (\ref{CoordinateTrnaformationRN2}), which change the action $\int Mdt$ of the black hole. 
The  coordinate transformation in Eq.(\ref{CoordinateTrnaformationRN2}) has two singularities (at two horizons). They produce the  change in the action, which imaginary part determines the macroscopic quantum tunneling from the PG RN black hole to its static partner:
\begin{eqnarray}
 P_{\text{BH-RN}\rightarrow \text{static-RN}}  =  \exp{ \left(-2\,{\rm Im} \int M  t \right)} =
 \label{P_RN1}
 \\
 = \exp{ \left(-2\,{\rm Im} \int M(d\tilde t  + f(r)dr ) \right)} = 
 \label{P_RN2}
 \\
 = \exp{ \left(-2M\,{\rm Im}  \int f(r)dr  \right)} = 
 \label{P_RN3}
 \\
 = \exp{ \left(-\pi (r_+ + r_-)^2/G\right)} =   \exp{ \left(-4\pi M^2G\right)} \,.
\label{P_RN4}
 \end{eqnarray}
The probability of the tunneling appears to be the same as for the neutral black hole with the same mass $M$. 

The quantum tunneling between the PG black hole and its static partner can be considered as the quantum fluctuation, which is determined by the difference in the entropy of the two objects.\cite{Landau_Lifshitz} Then  under the natural assumption that the entropy of the intermediate fully static object is zero, the Eq.(\ref{P_RN4}) gives the entropy of the RN black hole in agreement with  Eq.(\ref{entropyRNtilde}):
\begin{equation}
S_\text{BH-RN}(M,Q)=S_\text{BH}(M,0) =4\pi M^2 \,.
\label{entropy_RN}
\end{equation}
The  consideration of the further tunneling -- from the static RN state to the RN white hole state -- provides the negative entropy for the Reissner-Nordstr\"om white hole,
 \begin{equation}
S_\text{WH-RN}(M,Q)=- S_\text{BH-RN}(M,Q)= -4\pi M^2
 \,.
\label{entropy_RN2}
\end{equation}

All this demonstrates that the factor $(1+Q^2/r^2)$ in Eqs.(\ref{PG_RN2}) and (\ref{velocity_RN2})  is irrelevant for the entropy and Hawking temperature of the RN black hole. This suggests that this factor can be removed by some adiabatic process. We consider the possibility of such process in the following section \ref{Adiabatic}.

 \subsection{Entropy from adiabatic transformation}
\label{Adiabatic}

The independence of $\tilde S_{\rm RN}$ and $\tilde T_{\rm H}$ on the charge $Q$ may suggest that the RN black hole  can be obtained from the neutral BH by adiabatic process, at which the entropy does not change. But this would contradict to the conservation of charge $Q$ in the adiabatic process. However, let us remember that we used the equations, in which the fine structure constant 
$\alpha$ was hidden. With the parameter $\alpha$ involved one has the following dependence on $\alpha$:
 \begin{equation}
r_+r_-= \alpha Q^2\,\,, \,\,  r_\pm =  M \pm  \sqrt{M^2 - \alpha Q^2} \,,
\label{entropyRNalpha}
\end{equation}
Here $Q$ is conserved number, when measured in terms of the electric charge of electron, while the parameter $\alpha$ -- the fine structure constant -- is the property of the quantum vacuum, which can be varied by changing the parameters of the Standard Model. 
The gravity with varying $\alpha$  is the particular case of gravity with nonlinear electrodynamics (discussion of the latter see e.g. in Refs. \cite{Odintsov2017b,Kruglov2020,Amirabi2021} and references therein).

Now we can adiabatically transform the RN black hole  by slow change of this parameter $\alpha$ to zero at fixed mass $M$ and fixed quantum number $Q$.
Since in the adiabatic process the entropy does not change, it is the same as for $\alpha=0$, and thus it does not depend on $Q$ giving rise to Eq.(\ref{entropyRNtilde}):
 \begin{equation}
\tilde S_{\rm RN}(\alpha >0,M)= \tilde S_{\rm RN}(\alpha =0,M)=4\pi M^2 \,.
\label{entropyRNtildePositive}
\end{equation}

At first glance, it looks impossible to fix all three quantities (mass $M$, charge $Q$ and entropy $S$).
However, the RN black hole has extra degrees of freedom in the thermodynamics, which come from the inner horizon. 
When $\alpha$ slowly decreases to zero,  the two horizons in Eq.(\ref{entropyRNalpha}) move slowly  with conservation of the charge number $Q$ and energy $M$ until the inner horizon disappears.  In such slow process the entropy does not change, and thus it is the same as the entropy of the neutral black hole without inner horizon. 

It is interesting to consider the extension to the negative $\alpha$, where only single horizon remains.

 \subsection{Negative $\alpha$ and black holes with zero mass}

Let us consider electrodynamics with the inverse sign of the fine structure constant $\alpha=-|\alpha|$, and ignore the vacuum instability.  
For simplicity we consider the black hole with zero mass, $M=0$, which is now possible.
Then Eq.(\ref{r+r-}) becomes:
  \begin{equation}
  r_-= -r_+  \,\,, \,\, r_+^2=-  r_+r_-= |\alpha|Q^2  \,.
\label{r+r-BTZ_RN}
\end{equation}
One has  $r_-<0$, which means that the inner horizon is absent.

The metric is
\begin{eqnarray}
ds^2= - dt^2\frac{r^2 -|\alpha|Q^2}{r^2} + dr^2\frac{r^2}{r^2 -|\alpha|Q^2} +r^2 d\Omega^2 =
\label{rmetricBTZ_RN1}
\\
= - dt^2\left(1- \frac{|\alpha|Q^2}{r^2} \right)+ dr^2\frac{1}{\left(1- \frac{ |\alpha|Q^2}{r^2} \right)} +r^2 d\Omega^2.
\label{rmetricBTZ_RN2}
\end{eqnarray}
This looks as the 3+1 analog of the  2+1 Banados-Teitelboim-Zanelli (BTZ) black hole\cite{Banados1992} with mass $m=|\alpha|Q^2/8$, but without ADS at infinity:
\begin{equation}
ds_{\rm BTZ}^2=-N^2dt^2 + \frac{dr^2}{N^2} + r^2 d\phi^2
\,,
\label{BTZMetric}
\end{equation}
\begin{equation}
N^2 = r^2 - 8m\,.
\label{N2}
\end{equation}

Since the inner horizon is absent, the Painleve-Gullstrand  form is applicable:
\begin{equation}
ds^2= - dt^2(1-{\bf v}^2) - 2dt\, d{\bf r}\cdot {\bf v} + d{\bf r}^2 \,.
\label{PGmetric2}
\end{equation}
It is determined everywhere with shift velocity
\begin{equation}
v^2= \frac{ |\alpha|Q^2}{r^2}  \,,
\label{v}
\end{equation}
where different signs of the shift velocity correspond to the black and white holes.
For the nonzero black hole mass one has:
\begin{equation}
v^2=  \frac{2M}{r}  + \frac{|\alpha|Q^2}{r^2}  \,,
\label{v_general}
\end{equation}

The entropy of the black and white hole can be obtained in particular by the method of coordinate transformations between the black, white and fully static objects discussed in Sec.\ref{coordinate}:
\begin{equation}
dt \rightarrow dt \pm dr \frac{v}{1-v^2} \,.
\label{CoordinateTransformation}
\end{equation}
The probability of the macroscopic quantum tunneling from the black hole to its static partner with the same mass $M$ is given by the imaginary part in the action on this path:
\begin{eqnarray}
 P_{\text{BH}\rightarrow \text{static}} = \exp{ \left(-2{\rm Im} \int M dt \right)} = 
 \label{ImaginaryAction1}
 \\
 = \exp{ \left(-2{\rm Im} \int dr \frac{Mv}{1-v^2}  \right)} \,.
 \label{ImaginaryAction2}
 \end{eqnarray}
It is clear that for $M=0$ the probability $P_{\text{BH}\rightarrow \text{static}}=1$.
This means that for $\alpha<0$ the black hole with zero mass and their static and white partners are the same objects,
and thus their entropy is zero, $\tilde S_{\rm RN}(M=0,Q)=0$. This is in agreement with Eq.(\ref{entropy_RN2}), where the entropy of the Reissner-Nordstr\"om black hole depends only on its mass $M$, and does not depend on its charge $Q$ .

  \subsection{Discussion IV}

We calculated the  entropy of Reissner-Nordstr\"om black hole and the corresponding temperature of the thermal Hawking radiation
using several different approaches. These are: 

(i) The method of semiclassical tunneling, which is used for calculation of the Hawking temperature in Sec.\ref{HawkingRadiation}. 

(ii) The cotunneling  mechanism -- the coherent sequence of tunneling at two horizons, each determined by the corresponding Hawking temperature in Sec. \ref{Interpretation}. 

(iii) The calculation of the  macroscopic quantum tunneling from the RN black hole to the RN white hole using the method of singular coordinate transformations in Sec.\ref{Singular}. 

(iv) The adiabatic change of the fine structure constant $\alpha$ to zero and thus the adiabatic transformation from the RN black hole to the Schwarzschild black hole in Sec.\ref{Adiabatic}. 

All of the approaches give the same result. The correlations between the inner and outer horizons
lead to the total entropy and to the temperature of Hawking radiation, which depend only on mass $M$ of the black hole and do not depend on the black hole charge $Q$.

 The full agreement confirms the validity of different procedures used in this paper. In particular, this demonstrates that some singular coordinate transformations violate the general covariance in general relativity: they transform  the initial state to the physically (thermodynamically) different state. This corresponds to the spontaneously broken symmetry with respect to the general coordinate transformations, which leads to the existence of the non-equivalent degenerate states (black hole and white hole). While the physical laws are invariant under the singular coordinate transformations, the degenerate states are not invariant under these transformations and  are transformed to each other.
 
The presented approach also supports the statement  that the (anti)symmetry between the black and white holes can be extended to their entropy and temperature.
The Schwarzschild black hole and the Schwarzschild white whole are described by the metrics with opposite shift vectors. The shift vector changes sign under time reversal, which transforms the BH to WH. The absence of the time reversal invariance for each of these holes makes these states non static, but still the metric is stationary (time independent), and thus the entropy and temperature can be well defined.
The Schwarzschild BH and Schwarzschild WH have the opposite entropies  $S_\text{WH}(M)=- S_\text{BH}(M)$ and the opposite Hawking temperatures, $T_\text{WH}(M)=- T_\text{BH}(M)$. For the intermediate static hole the time reversal symmetry is not violated, and this object has zero temperature and zero entropy, $S_\text{static} =T_\text{static} =0$.

This approach has been extended to the Reissner–Nordstr\"om black hole with two horizons and the similar results were  obtained: $S_\text{WH}(Q,M)=- S_\text{BH}(Q,M)$ and  $S_\text{static}=0$. Moreover, the entropy of RN black hole appeared to be the same as that of the Schwarzschild BH with the same mass $M$, 
$S_\text{BH}(Q,M)=S_\text{BH}(Q=0,M)$, see Eq.(\ref{entropy_RN}). This deviates from the area law, which can be ascribed to the contribution of both horizons to entropy.  That is why the quantum tunneling process of the splitting of the  Reissner–Nordstr\"om black hole into two black holes gives the same result in Eq.(\ref{tunnelingM1M2}) as for Schwarzschild black hole.

It is interesting to consider the other objects including  the Kerr black and white holes, where time reversal symmetry is violated by rotation. In this case the coordinate transformations produces the singularity in action not only in $\int M\,dt$, but also in $\int J\,d\phi$, where $J$ is angular momentum and $\phi$ is the polar coordinate. The proper coordinate system can be found in Refs.\cite{Doran2000,Hamilton2008}. The recent discussion of the Painleve-Gullstrand forms and their extensions can be found in Refs. \cite{Faraoni2020,Visser2020}.
The entropy of the Kerr black was  obtained using the method of the adiabatic transformation.\cite{Volovik2021d} 
The result was similar to that for the RN black hole: the entropy depends only on the mass $M$ of the black hole:
$S_\text{BH}(M,J)=S_\text{BH}(M,J=0)$

The consideration can be extended to the other black holes with several horizons,\cite{Shankaranarayanan2003,Singha2021,Azarnia2021} in particular to the Reissner-Nordstr\"om-de Sitter black hole. In Ref. \cite{Singha2021} in the limit of $\Lambda=0$ the equations (\ref{HawkingTtilde1}) and (\ref{entropyRNtilde}) for the Reissner-Nordstr\"om black hole in Minkowski environment are reproduced.
However, it looks that de Sitter  environment   requires the special consideration. To see that let us consider the pure  de Sitter spacetime.

\section{de Sitter Universe}
\label{dS}

\subsection{de Sitter vacuum, cosmological constant problem and local temperature}

The issue of the stability of the de-Sitter vacuum is still an unsolved problem. This problem is related to the also unsolved problem of cosmological constant and dark energy. The particular solution of the cosmological constant problem is provided by the $q$-theory.\cite{KlinkhamerVolovik2008,KlinkhamerVolovik2021}  The $q$-theory describes the dynamics and thermodynamics of the  dark energy in terms of the 4-form field. This field was originally used by Hawking\cite{Hawking1984} for consideration of the vacuum energy. Note that the 4-form field is not the abstract quantity. For example, it may naturally enter the Diakonov theory of quantum gravity\cite{Diakonov2011} as the fundamental field.\cite{Volovik2021F}

In thermodynamics, the $q$-theory demonstrates, that in the equilibrium Minkowski vacuum there is the natural cancellation of the vacuum energy without any fine tuning. It is the consequence of the  Gibbs-Duhem relation, which does not depend on the microscopic (trans-Planckian) details and is applicable to nonrelativistic and relativistic  systems including the quantum vacuum. 

However, the problem remains in dynamics of vacuum energy, where the result of the relaxation of  the initial state with the large dark energy depends on the stability or instability of the de Sitter expansion. If the de Sitter expansion is stable and does not dissipate, then  both  the Minkowski and de Sitter vacua serve as the attractors in the vacuum dynamics.  To exclude the de Sitter attractor some mechanism of the decay of the de Sitter vacuum is necessary. In other words, while the $q$-theory solves the cosmological constant problem for the equilibrium vacuum, the decay of the de Sitter vacuum is the necessary condition for the dynamical solution of the cosmological constant problem within this theory.\cite{KlinkhamerVolovik2017} 

When studying the stability of the de-Sitter expansion, it is necessary to take into account the important differences between the properties of the ideal de Sitter expansion and that of the perturbed de Sitter, when it is deformed by matter fields; by the expansion history; or by other factors, which violate the original de Sitter symmetry. 

This symmetry is violated, for example, by an atom living in the background of de Sitter vacuum. The interaction of the atom with the de Sitter vacuum leads to ionization of the atom in the absence of electric field, the process which is not possible in the Minkowski vacuum due to energy conservation.  The ionization in the de Sitter vacuum can be again described by the quantum tunneling of electron from atom to outside, which is possible in the de Sitter vacuum. 

 \subsection{Double Hawking local temperature}
 \label{DoubleSec}

The corresponding stationary but not static PG metric is given by Eq.(\ref{PGmetric}) with the shift velocity
 \begin{equation}
  {\bf v}( {\bf r})=\pm H {\bf r}
\,,
\label{ShiftStatic}
\end{equation}
where $H$ is the Hubble parameter in the de Sitter universe. Using this metric, which does not contain singularities, one can calculate the  rate of ionization of atoms in de Sitter space in the semiclassical tunneling approach. The  ionization  looks as thermal, but with the effective temperature, which  is twice the traditional Hawking temperature $T_{\rm H}$ describing radiation from the cosmological event horizon:\cite{Volovik2009b} 
\begin{equation}
 T_{\rm loc}=2T_{\rm H} \,\,,  \,\, T_{\rm H} = \frac{H}{2\pi} \,.
\label{localT}
\end{equation}

Eq.(\ref{localT}) is valid in the regime, when  the relevant classical trajectories of electron in the ionization process are well within the event horizon. This means that effective temperature is the local temperature, and thus this process of ionization has no relation to the cosmological horizon.  The same local temperature describes the process of  the splitting of the composite particle with mass $m$ into two components with $m_1 +m_2>m$, which is also not allowed in the Minkowski vacuum.\cite{Bros2008,Bros2010,Volovik2009b}  The probablity of this process  (for $m\gg T_{\rm H}$):
\begin{eqnarray}
\Gamma(m \rightarrow m_1+m_2) \sim \exp{\left(-\frac{\pi (m_1+m_2-m)}{H}\right)} =
\nonumber
\\=  \exp{\left(-\frac{m_1+m_2-m}{ T_{\rm loc}}\right)}\,.
\label{mto2}
\end{eqnarray}

If only single scalar field with mass $m$ is present, the process of the splitting of a particle to two particles becomes the process of duplication of particles, i.e. the particle transforms to  two particles of the same bosonic field, $m \rightarrow 2m$. The equation  (19) in Ref.\cite{Volovik2009b} gives 
for $m_1=m_2=m$ the probability of this process
\begin{equation}
\Gamma(1 \rightarrow 2) \sim \exp{\left(-\frac{\pi (2m-m)}{H}\right)}= \exp{\left(-\frac{ m}{T_{\rm loc}} \right)} \,.
\label{mto2m}
\end{equation}
This coincides with the decay rate of massive field in the de Sitter Universe $\Gamma(1\rightarrow 0) \sim  \exp{\left(-\frac{m}{T_{\rm loc}} \right)}$ obtained in Ref. \cite{Jatkar2012}. 
This factor $\exp{\left(-\frac{m}{2T_{\rm H}} \right)}$  in the decay rate of massive field is discussed in detail in Ref. \cite{Maldacena2015} devoted to the so-called Cosmological Collider (inflation viewed as a particle accelerator).
It is stressed that this factor differs from conventional thermal factor  $\exp{\left(-\frac{m}{T_{\rm H}}\right)}$, which characterizes the creation of a pair of massive particles.
 
 Though it looks that the Hawking radiation is less important than the local processes, its consideration allows us to determine the local temperature in terms of the Hawking temperature, $T_{\rm loc}=2T_{\rm H}$.
Eq. (\ref{mto2m}) demonstrates that creation of new particles  is compensated  by their decay,
 $n\Gamma(1\rightarrow 2)=n\Gamma(1\rightarrow 0)$, where $n$ is particle density. This detailed balance, however, does not determine  the distribution function $n$ in equilibrium. The equilibrium particle density can be found from the consideration of Hawking radiation from de Sitter cosmological  horizon.
 
In the Hawking radiation, two particles are created: one is inside horizon and the other is outside. For the local observer at the cosmological horizon the Gibbons-Hawking radiation   is observed as the coherent creation of two particles with mass $m$. Thermal excitation of two particles corresponds the environment with temperature, which is double Hawking temperature, $T_{\rm loc}=2T_{\rm H}$, i.e. 
$\Gamma(0 \rightarrow 2) = \exp{\left(-\frac{m}{T_{\rm H}} \right)}=  \exp{\left(-\frac{2m}{T_{\rm loc}} \right)}$. Since the de Sitter state is symmetric, all the points in its space are equivalent, and thus at each point the Hawking radiation is viewed as the creation of pairs from the vacuum.
The detailed balance $\Gamma(0 \rightarrow 2) =n\Gamma(1\rightarrow 2)= n \Gamma(1 \rightarrow 0)$ gives $n \sim  \exp{\left(-\frac{m}{T_{\rm loc}}\right)}= \exp{\left(-\frac{m}{2T_{\rm H}} \right)}$.

Note that double Hawking temperature also appeared in some calculations of the black hole radiation. However, this was the result of using the metric with singularities at the horizon. The detailed discussion is in Ref.\cite{Volovik2022s}.

\subsection{Global vs local temperature}
 \label{GlobalLocalSec}

Let us consider the $2T_{\rm H}$  problem using Klein–Gordon equation  for massive field
in a curved background,\cite{Akhmedov2006} 
which leads to the relativistic Hamilton--Jacobi equation for the classical action:
 \begin{equation}
g^{\mu\nu}\partial_\mu S \partial_\nu S + m^2 =0\,.
\label{HJ}
\end{equation}
In  the PG metric (\ref{PGmetric}), which does not have coordinate singularity at the horizon, one has for the fixed energy $E$:\cite{Akhmedov2006} 
 \begin{equation}
-E^2 +(1-v^2) \left(\frac{dS}{dr}\right)^2 + 2vE\frac{dS}{dr}+m^2=0\,.
\label{KG}
\end{equation}
This gives for the classical action
 \begin{equation}
S=-\int dr \frac{Ev}{1-v^2} \pm \int  \frac{dr}{1-v^2}\sqrt{E^2-m^2(1-v^2)} \,.
\label{SPG}
\end{equation}

Let us first consider the tunneling events related to the cosmological horizon.
For the plus sign in the second term the imaginary parts of two terms cancel each other. This corresponds to the incoming trajectory of the particle.
For the minus sign in the second term the total tunneling exponent describes the radiation with the Hawking temperature $T_{\rm H}$:
\begin{eqnarray}
 \exp{ \left(-2\,{\rm Im}\, S\right)}= \exp{ \left(-\frac{E}{2T_{\rm H}} \right)} \exp{ \left(-\frac{E}{2T_{\rm H}} \right)} =
 \nonumber
 \\
 =\exp{ \left(-\frac{E}{T_{\rm H}} \right)} \,.
\label{TH2}
\end{eqnarray}
Note that the above calculations do not depend on the type of the horizon, and are applicable also to the black hole horizon.

Each of the two terms in Eq.(\ref{SPG}) corresponds to the effective temperature $2T_{\rm H}$. 
That is why if one of the two terms is lost in calculations, the temperature $2T_{\rm H}$ erroneously emerges, as this happens in some calculations of the Hawking radiation from black holes. But for the de Sitter case the temperature $2T_{\rm H}$ is physical. 

Till now we considered  the imaginary part the action (\ref{SPG}) for particle with energy $E>m$ on the trajectory  in the complex plane  which connects the trajectory inside the horizon and the trajectory outside the horizon. This corresponds to the creation of two particles: one inside the horizon and another one outside the horizon. However, there is also the trajectory which allows for the creation of a single particle fully inside the cosmological horizon.\cite{Volovik2022s}  In this creation from "nothing", the particle with mass $m$ must have zero energy, $E=0$. This is possible, as follows from the second term in Eq.(\ref{SPG}), which  gives the following imaginary part of the action at $E=0$:
\begin{eqnarray}
 {\rm Im}\, S(E=0)= m \int _0^{1/H}\frac{dr}{\sqrt{1- r^2H^2}} =\frac{\pi}{2}\frac{m}{H}\,.
 \label{LocalCreation1}
 \end{eqnarray}
 The probability of radiation
 \begin{eqnarray}
  \exp{ \left(-2\,{\rm Im}\, S\right)}= \exp{ \left(-\frac{\pi m}{H}\right)}=\exp{ \left(-\frac{m}{2T_{\rm H}}\right)}\,,
   \label{LocalCreation2}
\end{eqnarray}
 corresponds to the thermal creation of particles by the environment with local  temperature equals to the double Hawking temperature. It is the same local temperature, which characterizes the thermal ionization of atoms in the de Sitter space.

The processes of single and double particle creation take place also in the black hole physics. Let us consider the radiation measured by the observer at finite distance $R_{\rm o} \gg R$ from the black hole:\cite{Jannes2011,Volovik2022s} 
\begin{equation}
 p(E) \sim  \exp\left(-\frac{E}{T_{\rm H}}\right) \exp\left(-2R_{\rm o}\theta(m-E)\sqrt{m^2-E^2}\right).
 \label{SingleBH3}
\end{equation} 
For $E<m$, there are two tunneling exponents. The first exponent comes from the horizon and corresponds to the conventional Hawking radiation. The second one describes the process of tunneling of the created particle from the horizon to detector.

 The probability $p(E=0)$ describes the creation of single particle outside the horizon, i.e. without the entangled partner. This is similar to the process, which we discussed for the case of the de Sitter expansion in Eq.(\ref{LocalCreation2}). As distinct from the de Sitter Universe, in case of the black hole this process is exponentially suppressed and is not described by the fundamental temperature. This demonstrates that the de Sitter expansion with its local temperature $2T_{\rm H}$ is the very specific spacetime.

  \subsection{Local thermodynamics in de Sitter Universe}
 \label{ThermodynamicSec}

Let us consider how this local temperature $T_{\rm loc}=2T_{\rm H}$ may enter statistical physics of the de Sitter Universe.\cite{Volovik2021g} 
According to Ref. \cite{Podolskiy2018}  the probability of fluctuation of the potential of the scalar field $\Phi$ in the de Sitter spacetime is
\begin{equation}
P \propto \exp{ \left(- \frac {8\pi^2}{3H^4} \delta V(\Phi) \right)} =\exp{ \left(- \frac {F(\Phi)}{T} \right)}\,,
\label{fluctuations}
\end{equation}
where $F$ is the free energy of the matter field. 

There are two ways to interpret this probability. The traditional interpretation is in terms of the Hawking temperature: \cite{Podolskiy2018}
\begin{equation}
 T=T_{\rm H}=\frac{H}{2\pi}\,\,\,,\,  F(\Phi)=\frac{4\pi}{3H^3} \delta V(\Phi) \,.
\label{interpretation1}
\end{equation}
Here the free energy is determined in the region within the cosmological horizon with volume  $\frac{4\pi}{3H^3}$, i.e.  the volume of a given Hubble patch. 

On the other hand, Eq.(\ref{fluctuations}) describes the fluctuations of the matter fields on the background of the de-Sitter vacuum. That is why they should be discussed in terms of the physical temperature, i.e. the local temperature experienced by the matter fields, $T=T_{\rm loc}$. 
Then  from Eq.(\ref{fluctuations}) one obtains 
\begin{equation}
 T=T_{\rm loc}=\frac{H}{\pi}\,\,\,,\,  F(\Phi)=\frac{8\pi}{3H^3} \delta V(\Phi) \,.
\label{interpretation2}
\end{equation}
Now the free energy is determined in the whole volume of the expanding patch.
The static coordinates with Hubble volume $\frac{4\pi}{3H^3}$ inside the horizon cover only half of the expanding Poincare patch. The extension to the whole expanding Poincare patch requires doubling  of the effective volume, and this is the origin of the volume $\frac{8\pi}{3H^3}$  in Eq.(\ref{interpretation2}). This interpretation is in terms of the local physics and does not depend on the existence of the event horizon, which artificially separates the patches inside and outside the horizon. 

Note that the tunneling process of the particle decay, which is characterized by the effective temperature $T=T_{\rm loc}$, is not possible in the pure de Sitter spacetime, since the particle is simply absent in the pure de Sitter. Both the decaying particle  and the observer violate the symmetry of the pure de Sitter spacetime, which makes the process possible. 

So the thermodynamics of fluctuations on the background of the de Sitter is fully determined by the local temperature, which is not related to the cosmological horizon. The reason, why this local temperature is twice larger than the Hawking temperature, follows from the geometry  of the de Sitter space.

\subsection{Entropy and temperature of de Sitter Universe from singular coordinate transformations}

 Let us now apply the procedure of singular coordinate transformations in Sec. \ref{coordinate} for the calculations of the de Sitter entropy.  The corresponding fully static metric, which is time reversal symmetric, is:
\begin{equation}
ds^2=- \left( 1- H^2r^2\right)dt^2 + \frac{dr^2}{1- H^2r^2} + r^2 d\Omega^2
\,.
\label{dSStaticMetric}
\end{equation}
The coordinate transformation between the expanding and static metrics is given by Eq.(\ref{CoordinateTransformation}).
The probability of the macroscopic quantum tunneling from the de Sitter to its static partner with the same energy $E$ is given by the imaginary part in the action:
\begin{eqnarray}
 P_{\text{dS}\rightarrow \text{static}} = \exp{ \left(-2{\rm Im} \int E dt \right)} = 
 \label{ImaginaryAction1dS}
 \\
 = \exp{ \left(-2{\rm Im} \int dr \frac{Ev}{1-v^2}  \right)} =
 \label{ImaginaryAction2dS}
  \\
 = \exp{ \left(-2{\rm Im} \int dr \frac{EHr}{1-H^2r^2}  \right)} =
 \label{ImaginaryAction3dS}
 \\
 =  \exp{ \left(-\frac{\pi E}{H}\right)} \,.
\label{ImaginaryActiondS}
\end{eqnarray}

From the connection between the tunneling transition and the difference in the entropy of the initial and final states it follows that  the entropy of the expanding Universe is 
\begin{equation}
S_\text{dS}= \frac{\pi E}{H}
\,.
\label{dSentropy}
\end{equation}
This equation allows us to study thermodynamics of the de Sitter Universe
The main problem is that the proper energy $E$ of the de Sitter Universe is not well determined, and there are different suggestions such as $E=1/(2H)$,  $E=-1/(2H)$ and  $E=-1/H$.\cite{Padmanabhan2002}

Let us first first consider the "natural" choice that the entropy has the conventional form as the quarter of the  area of the cosmological horizon, $S_\text{dS}=A/4=\pi/H^2$. Then from Eq.(\ref{dSentropy})  one obtains:
\begin{equation}
E_\text{dS}=\frac{1}{H} = \frac{8\pi}{3H^3}\Lambda
\,,
\label{dSenergy}
\end{equation}
This corresponds to the vacuum energy in the whole volume of the expanding patch, $V= 8\pi/3H^3$.
This allows us to consider the temperature of the de Sitter Universe. If one determines temperature as simply the variation of energy over entropy, one obtains that the temperature of the de Sitter Universe coincides with the Hawking temperature of the cosmological horizon:
\begin{equation}
T_\text{dS}= \frac{dE}{dS}=\frac{H}{2\pi}=T_\text{H}
\,.
\label{dStemperature}
\end{equation}

However, all this is not so simple since we did not take into account the term $-pdV$, where $p$ is the vacuum pressure, which in the de Sitter vacuum is with minus sign the cosmological constant, $p=-\Lambda$.
To avoid the pressure term we can vary the entropy in Eq.(\ref{dSentropy}) over $dE$ at fixed $H$, which corresponds to the variation at fixed volume $V$ (this can be done by varying the Newton constant $G$ without  changing $H$).
Then one obtains 
\begin{equation}
T_\text{dS}= \frac{dE}{dS}=\frac{H}{\pi}=2T_\text{H}
\,.
\label{dStemperature2}
\end{equation}

  \subsection{Discussion V}
  
  In case of de Sitter spacetime, two different temperatures are important. One of them is the Hawking temperature  $T_{\rm H}$, which describes the creation of the entangled pair of particles. Another one is the local  temperature  $T_{\rm loc}=2T_{\rm H}$, which describes the thermal creation of a single (non-entangled) particle inside the cosmological horizon. The connection between the local and global temperatures is determined by the specific geometry of the de Sitter expansion. Which of these two temperature determines the thermodynamics of the de Sitter spacetime is an open question, and we considered the possibility that the thermodynamics is governed by the higher temperature,   $T_{\rm loc}=2T_{\rm H}$.

In this respect it would be interesting to study the objects, which combine the properties of the black hole with the properties of the de Sitter spacetime.
These compact objects (gravastars) are the black or white holes which have the de Sitter spacetime inside the horizon.\cite{Chapline2003,Dymnikova2020} In terms of the PG metric this object has the shift velocity in Eq. (\ref{velocity}) outside the horizon and 
 the shift velocity in Eq.(\ref{ShiftStatic}) inside the horizon with $H=1/R$,\cite{Volovik2021k}  where the sign $-$ for the black gravastar and the sign $+$ for the white gravastar. Both do nor experience the Hawking radiation.
 In principle, such black hole objects can be developed in the process of a black hole's evaporation, or even due to the rapidly developed  instability inside the horizon. In the latter case the characteristic time scale of relaxation to the equilibrium state is $\tau \propto R/c\propto 1/T_{\rm H}$. This corresponds to the Bekenstein-Hod universal bound,\cite{Hod2007,Carullo2021} but now applied to the black hole with the de Sitter interior. On the other hand, the large white gravastar may simulate the Big Bang.\cite{Gibbs1998,Retter2012}
  
So, it looks that the modified de Sitter thermodynamics demonstrates that it can be described without consideration of the cosmological horizon and Hawking radiation.  The quantum tunneling process, which leads to the decay of the composite particle in the de Sitter vacuum, occurs fully inside the cosmological horizon and is fully determined by the local temperature, $T_{\rm loc}$. The free energy of the fluctuations of the matter fields also corresponds to the local temperature, $T_{\rm loc}$. It is not restricted by the region inside the horizon, i.e. it is also not related to the existence of the cosmological horizon. All this raises the question of the role of the cosmological horizon and Hawking temperature in the pure de Sitter vacuum. 

The  decay of the composite particles, which are excitations above the de Sitter vacuum,  does not directly lead to the decay of the vacuum itself. However, let us assume that  the de Sitter state contains the thermal matter with the local temperature $T_{\rm loc}$. Then the interaction between the thermal matter and the dark energy during the evolution of the Universe leads to the decay of the vacuum energy density $\rho_V$ and of the Hubble parameter $H$ according to the following power law:\cite{Volovik2021g}
 \begin{eqnarray}
H \sim E_{\rm Pl} \left( \frac{t_{\rm Pl}}{t}\right)^{1/3}  \,,
\label{DecayLawH}
\\
\rho_V \sim E_{\rm Pl}^4  \left( \frac{t_{\rm Pl}}{t}\right)^{2/3} \,.
\label{DecayLawV}
\end{eqnarray}
Here the Planck time $t_{\rm Pl}=G^{1/2}$ and Planck energy $E_{\rm Pl}=1/t_{\rm Pl}$ are introduced.
Such power law decay is discussed in different approaches. It is similar to that in Eq.(192) in Ref. \cite{Padmanabhan2003} (see also Ref.\cite{Gong2021}) and in Eq.(109) in Ref. \cite{Markkanen2018}. The  time scale of the decay of the de Sitter expansion, which follows from Eq.(\ref{DecayLawV}), $t_Q=E_{\rm Pl}^2/H^3$, corresponds to the time at which  de Sitter state looses coherence.\cite{Berezhiani2021}

On the other hand, the possibility of the decay of the pure de Sitter vacuum due to Hawking radiation remains unclear and requires the further consideration.\cite{Kamenshchik2021} This does not mean that the de Sitter vacuum is stable: this only means that the Hawking radiation alone does not lead to instability, i.e the de Sitter vacuum is stable with respect to
the decay via the Hawking radiation. The Hawking radiation does not lead to the change of the vacuum energy density, which generates the de Sitter expansion. This means that even if the pair creation takes place, the de Sitter expansion immediately dilutes the produced particles, and thus there is no vacuum decay in de Sitter. 
 
 There are many other mechanisms, not related to the Hawking radiation, which could lead to the decay of the de Sitter spacetime,\cite{Starobinskii1977,Starobinskii1979,Starobinskii1983,StarobinskyYokoyama1994,Polyakov2008,Polyakov2012,Akhmedov2014,Palti2019}
including the infrared instability, instability due to the dynamic effects of a certain type of quantum fields, instability towards spontaneous breaking of the symmetry of the de Sitter spacetime or the instability towards the first order phase transition in the vacuum, etc.
But in most cases either the de Sitter vacuum is not perfect, i.e. there are deviations from the exact de Sitter, and the de Sitter symmetry is lost, or the vacuum energy is fine-tuned, i.e. the cosmological constant problem is ignored. The de Sitter instability, which avoids fine tuning, but uses the special vector field in Dolgov scenario,\cite{Dolgov1997} is in Ref.  \cite{Emelyanov-Klinkhamer2012}.
 
The problem of the dynamical stability of the de Sitter vacuum is directly related to the cosmological constant problem. The $q$-theory\cite{KlinkhamerVolovik2008}   demonstrates the solution of the problem in thermodynamics: in the equilibrium Minkowski vacuum the cosmological constant is nullified due to thermodynamics. However, it remains unclear whether the de Sitter state relaxes to the equilibrium. This depends on the  stability of the de Sitter vacuum.  If the de Sitter attractor is not excluded in dynamics, then the only possibility to solve the dynamical cosmological problem within the $q$-theory is to assume that the Big Bang occurred in the part of the Universe, which is surrounded by the equilibrium environment.\cite{KlinkhamerSantillan2019} In this case, any perturbation of the vacuum energy by the Big Bang, even of the Planck scale order, will inevitably relax to the equilibrium Minkowski vacuum with zero cosmological constant.
This relaxation does not require any fine-tuning, since it is dictated by the equilibrium environment.

  \section{Conclusion}

In conclusion, the macroscopic quantum tunneling elaborated in the early works by S.V. Iordansky, A.M. Finkel'shtein and E.I. Rashba in Landau Institute allows us to study the similar processes in cosmology. The probability of the processes of macroscopic quantum tunneling of cosmological objects is extremely small. However, the theoretical consideration of these processes allow us to make conclusions on entropy and temperature of the cosmological objects, which are rather unexpected. 

  {\bf Acknowledgements}.  This work has been supported by the European Research Council (ERC) under the European Union's Horizon 2020 research and innovation programme (Grant Agreement No. 694248).

\end{document}